\documentclass [12pt]{article}
\usepackage{amsmath}
\usepackage{amssymb}
\usepackage{graphics}
\usepackage{graphicx}

=240mm

\begin{document}

\title{X-ray multi-energy radiography with
scintillator-photodiode detectors}
\author{S.V.~Naydenov, V.D.~Ryzhikov, B.V.~Grinyov, \\
E.K.~Lisetskaya, A.D.~Opolonin, D.N.~Kozin \\
{\it Institute for Single Crystals of NAS of Ukraine}, \\ {\it 60
Lenin ave., 61001 Kharkov, Ukraine} }

\date{\empty}
\maketitle

\begin{abstract}
For reconstruction of the spatial structure of
many-component objects, it is proposed to
use multi-radiography with detection of  X-ray 
by combined detector arrays using detectors of
scintillator-photodiode type. A theoretical model has been
developed of multi-energy radiography for thickness measurements
of multi-layered systems and systems with defects. Experimental
studies of  the sensitivity, output signal of various inspection systems based
on scintillators $ZnSe(Te)$ and  $CsI(Tl)$, and object image reconstruction (with organics 
and non-ogranics materials singled out)  have been carried out.
\end{abstract}

\textbf{1.} The radiographic method (with signal conversion to
digital form) is one of the main directions of modern
non-destructive testing (NDT) [1, 2]. As radiation sources, X-ray
tubes are generally used, with characteristic radiation energy
from tens to hundreds $keV$. When this radiation is absorbed in
the studied objects, the processes that dominate are photo effect
and Compton scattering. Linear attenuation coefficients $\mu (E)$
at medium radiation energies are well known for most substances
and materials. In introscopy of the objects of large thickness,
radioactive sources with higher energies of penetrating X-ray and
gamma-radiation can also be used.

Inspection and technical diagnostics (TD) are based on scanning
(linear or planar) and subsequent topography of the
three-dimensional structure of the object. It is often needed to
carry out quantitative analysis of the internal structure of
materials. When the geometry is complex, as well as for systems of
variable thickness, multi-layered, multiply connected or
multi-component structure, conventional NDT methods ("one-energy",
but non-monochromatic) could be insufficient. The use of more
informative and more complex tomographic methods is not always
possible due to technical or economical reasons. Important
progress can be achieved here in relationship with a multi-energy
approach. Radiographic monitoring with separate detection of
radiation (in different ranges of the energy spectrum) at several
selected energies can give additional information on the internal
structure of the studied object. A block diagram of such method is
presented in Fig.1.

\begin{figure}[ht]
\centering
\includegraphics[width=5.52in,height=1.43in]{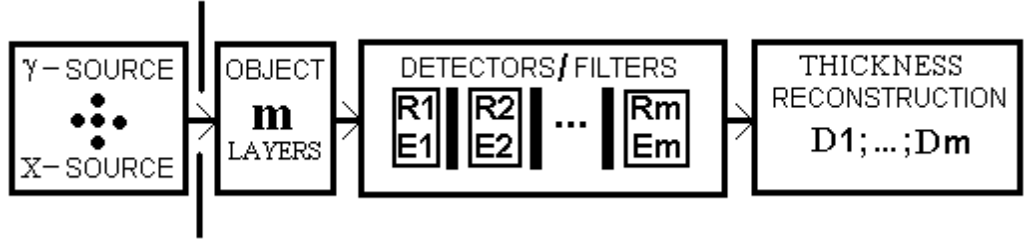}
\caption{Schematic diagram of multi-energy radiography of
thickness.} \label{fig:Fig1}
\end{figure}

\bigskip

\textbf{2.} In developing of the said aspect of multi-energy
radiography (MER), especially efficient are simple schemes of two-
and three-energy monitoring. Fig.2 shows characteristic cases of
mutual position of simple objects $A$, $B$ and $C$ as parts of a
``complex'' object or overlapping in the projection the defects A
and B on the main background $C$. In this case, carrying out of
structuroscopy (determination of thickness or defectoscopy) is
equivalent to solution of the inverse problems for 2- and 3-MER,
respectively. It follows from the theory that in the general case
the number of reconstructed thicknesses is the same as the
multiplicity of radiography, i.e., the number of separately
detected radiation ranges). Consequent local scanning of the
object allows us to reconstruct the profile of its internal
three-dimensional structure also in the case of variable
cross-section of the components that form it. To determine
thickness of separate components or size of inclusions, one has to
assume their chemical composition to be approximately known. This
refers to the two parameters that are principal for radiography --
effective atomic number $Z$ and density$\rho $ of each specific
material. Or, linear attenuation coefficients should be specified
for corresponding substances. For independent determination of
these $Z$ and $\rho $, it is also possible to use means of MER
[3].

\begin{figure}[h]
\centering
\includegraphics*[scale=0.6]{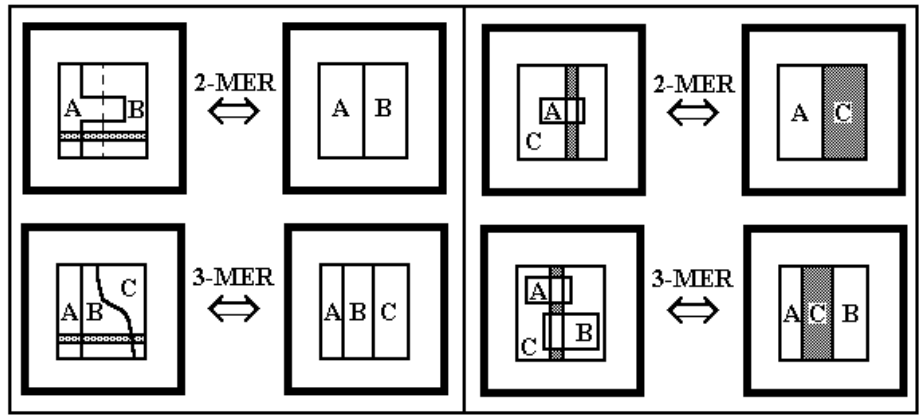}
\caption{Schematic diagram of reconstruction of three-dimensional
structure for many-component and multiply connected (i.e., with
defects) objects under linear scanning using 2- and 3-MER.}
\label{fig:Fig2}
\end{figure}

\bigskip

\textbf{3.} Theoretical model for thickness reconstruction by means of MER
uses the universal character of exponential attenuation of the quantized
radiation in monitoring objects and detectors. Passing over to logarithmic
(arbitrary) units of the detected signal normalized to the background value
(when the object is absent), radiography equations can be presented in a
simple form
\begin{equation}\label{eq1}
R_{i} = \sum\limits_{j = 1}^{M} \mu _{ij} D_{j} \; \quad ;
\end{equation}

\begin{equation}
\label{eq2} \mu _{ij} = \rho _{j} {\left[ {\tau \left( {E_{i}}
\right)Z_{j}^{4} + \sigma \left( {E_{i}}  \right)Z_{j} + \chi
\left( {E_{i}}  \right)Z_{j}^{2} } \right]}\;;\quad i = 1,\ldots
,m;\;j = 1,\ldots ,M \quad ,
\end{equation}
where $R\left( {E_{i}}  \right) \equiv R_{i} $ are reflexes
(registration data) at corresponding maximum absorption energies
within each monitoring range. Unknown are thicknesses $D_{i} $.
Matrix $\mu _{ij} $ (of linear attenuation coefficients) will be
specified, with energy dependencies on photo-effect $\tau $,
Compton scattering $\sigma $ è and pair generation effect $\chi $.
In the medium energy range up to $0.5\,MeV$, the latter scattering
channel can be neglected. Solving the linear system is the inverse
problem of MER. To obtain its univalent solution and to determine
the thicknesses, the number of layers $m$ should correspond to the
order $M$ of multi-energeticity, $m = M$. The general solution has
the form
\begin{equation}
\label{eq3} D_{i} = {\sum\limits_{j = 1}^{m} {\mu _{ij}^{ - 1}
R_{j}} }\;;\quad \det \mu _{ij} \ne 0 \quad ,
\end{equation}
where $\mu _{ij}^{ - 1} $ is the inverse matrix. In the case of
2-MER, it is convenient to write down explicit formulas
\begin{equation}
\label{eq4} D_{1} = {\frac{{\mu _{22} R{\kern 1pt} _{1} - \mu
_{12} R_{2} }}{{\mu _{11} \mu _{22} - \mu _{12} \mu _{21}}
}}\;;\quad D_{2} = {\frac{{\mu _{11} R_{2} - \mu _{21} R_{1}}
}{{\mu _{11} \mu _{22} - \mu _{12} \mu _{21}} }} \quad .
\end{equation}
We do not present here the somewhat clumsy expressions for 3-MER
case. In the general case, for determination of $D_{i} $ it is
necessary and sufficient that determinant $\det \mu _{ij} \ne 0$.
This implies a physical condition for MER feasibility:
\begin{equation}
\label{eq5} \forall i \ne j\quad \Rightarrow \quad \;{\left|
{E_{i} - E_{j}}  \right|} \gg \delta E_{noise} \quad ,
\end{equation}
where $\delta E_{noise} $ is the total noise level in the system
expressed in energy units. For one-energy radiography, separation
of the reconstructed ``images'' of the composed object by scanning
at one camera angle is not possible. This experimentally and
theoretically proven fact corresponds to the uncertainty of
expressions (\ref{eq4}) when their denominator becomes zero at
$E_{1} = E_{2} $.

\bigskip

\textbf{4.} For practical developments of MER, an important factor
is detector sensitivity of the inspecting system. In the Concern
``Institute for Single Crystals'', combined detectors of
``scintillator-photodiode'' type have been developed, which are
characterized by improved sensitivity (contrast and detecting).
The two-energy system was realized on the basis of a ``sandwich''
structure comprising two detectors of ``scintillator-photodiode''
type. It includes a low-energy detector (LED) based on
\textit{ZnSe(Te)} and a high-energy detector (HED) based on
\textit{CsI(Tl)}. Both theoretical calculations and experiments
show that such combination is the most efficient for
multi-radiographic inspection. In experiments on determination of
the detector sensitivity $S = {dI} \left/ {dD} \right. $ (Fig.3)
and output signal $I$ (Fig.4), the X-ray source used had anode
voltage $U_{a} = 40 \div 180\,kV$ and current $I_{a} = 0.4\,mA$ (a
tube with a $W$-shaped anode). Sensitivity $S$ and output signal
$I\left( {E} \right) \propto \exp {\left[ { - R\left( {E} \right)}
\right]}$ was determined in arbitrary units.

In choosing detectors, the following features were accounted for.
Scintillator \textit{ZnSe(Te)} has relatively small atomic number
$Z_{eff} = 32$, but its density is high enough to ensure efficient
absorption of the ionizing radiation in the low energy region.
Light output of \textit{ZnSe(Te)} can reach 100-130\% with respect
to \textit{CsI(Tl)} at absorbing thickness of $0.1 - 1\,mm$. As a
result, all this ensures substantial advantages of zinc selenide
for radiation detection in the $20 \div 80\,keV$ range as compared
with other scintillators and good filtration of this part of the
X-ray radiation spectrum. Our calculations have also shown that
optimum thickness values of scintillators for the two-energy
radiograph with $U_{a} = 140\,kV$ are: for LED \textit{ZnSe(Te)}
--  $0.6\,mm$; for HED \textit{CsI(Tl)}  --  $4\,mm$.

Fig.3 and Fig.4 show results of our measurements of the relative detecting
sensitivity and the output signal (reflex) for combined scintielectronic
detector arrays of different types. The data obtained confirm advantages of
the chosen type and design of the 2-energy inspection system. This physical
configuration has been realized in the Polyscan-4 two-energy introscope [4].
Images of a multi-component object obtained using this inspection system are
shown in Fig.5. This system also allows distinction between images
corresponding to materials with high and low atomic number, e.g., to detect
organic materials against the background of inorganics.

\begin{figure}[ht]
\centering
\includegraphics[width=4.98in,height=3.45in]{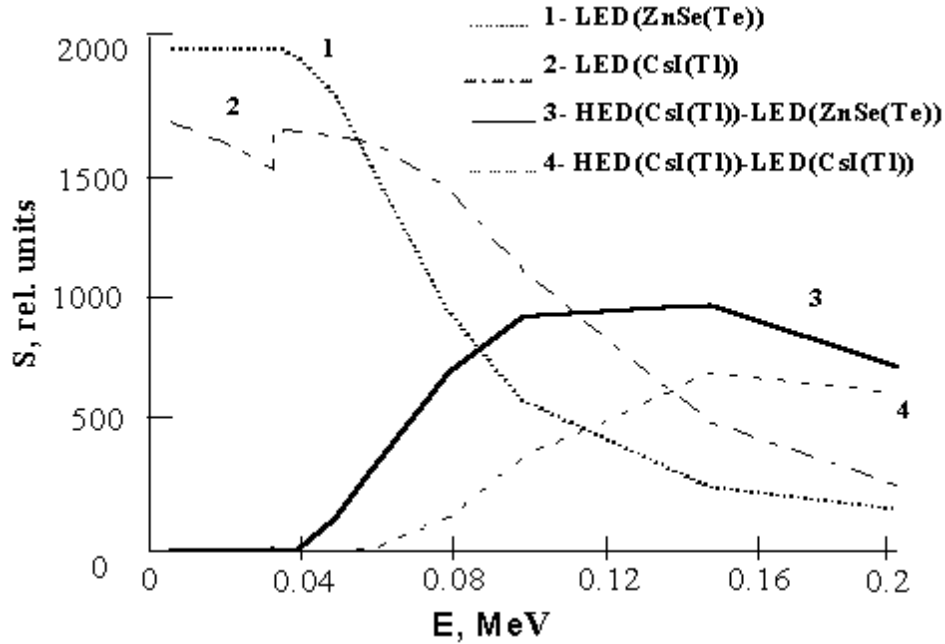}
\caption{Energy dependence of the detector sensitivity in a
two-level system with various combinations of scintillators.}
\label{fig:Fig3}
\end{figure}

\bigskip

\begin{figure}[ht]
\centering
\includegraphics[width=4.93in,height=3.42in]{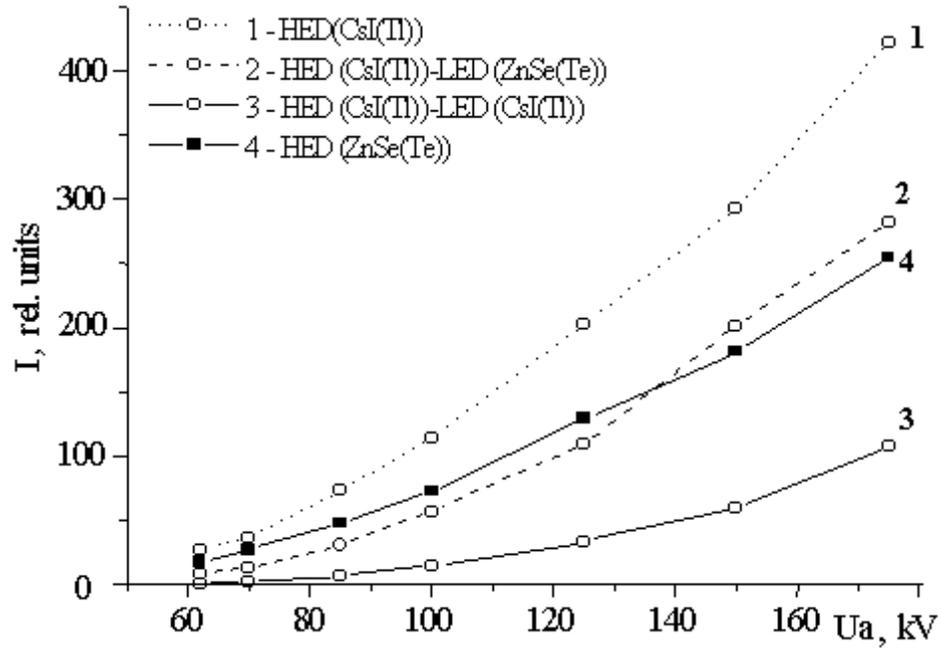}
\caption{Output signal of HED with LED active filter and various
combinations of scintillators as function of the tube
voltage.}\label{fig:Fig4}
\end{figure}

\bigskip

\textbf{5.} The developed scheme of multi-radiography can be
directly used for different control evaluations, especially in
topography of several surimposed ``layers''(or defects) or when
analysis under different camera angles is impossible. Quantitative
determination of thicknesses in a many-component structure makes
it possible to physically discern between physically surimposed
parts of one and the same piece or object. This substantially
increases contrast sensitivity of MER as compared with
conventional methods, which is important not only for technology,
but also for medical applications (separate diagnostics of soft
and bone tissues). Therefore, the proposed radiographic method of
multi-energy reconstruction of geometrical structure of the
objects can be useful for many applications in the field of NDT
and TD. This conclusion is also supported by the already achieved
positive results in industrial production of 2- and 3-energy
detectors of different types and modifications, e.g., [5, 6].

\bigskip

[1] 15$^{{\rm t}{\rm h}}$ Word Conference on NDT, Rome (Italy),
15-21 Oct., 2000, Abstracts Book, 800 p.

[2] R.M. Harrison, Nucl. Instr. and Meth. \textbf{A310}, pp. 24-34
(1991).

[3] S.V.~Naydenov, V.D.~Ryzhikov, \textit{Technical Physics
Letters}, vol. \textbf{28}, \# 5, pp. 357-360 (2002).

[4] The X-Ray Introscopy System of Luggage Customs Control
"Poliscan-4", Prospects, developed by STC RI \& SCB "Polisvit" PO
"Kommunar"; e-mail: stcri@isc.kharkov.com .

[5] Rapiscan Prospects. USA. - 2002, http://www.rapiscan.com .

[6] Heimann Prospects. Germany. - 2002,
http://www.heimannsystems.com .

\newpage

\begin{figure}[ht]
\begin{center}
\textbf{a)}
\includegraphics*[scale=0.6]{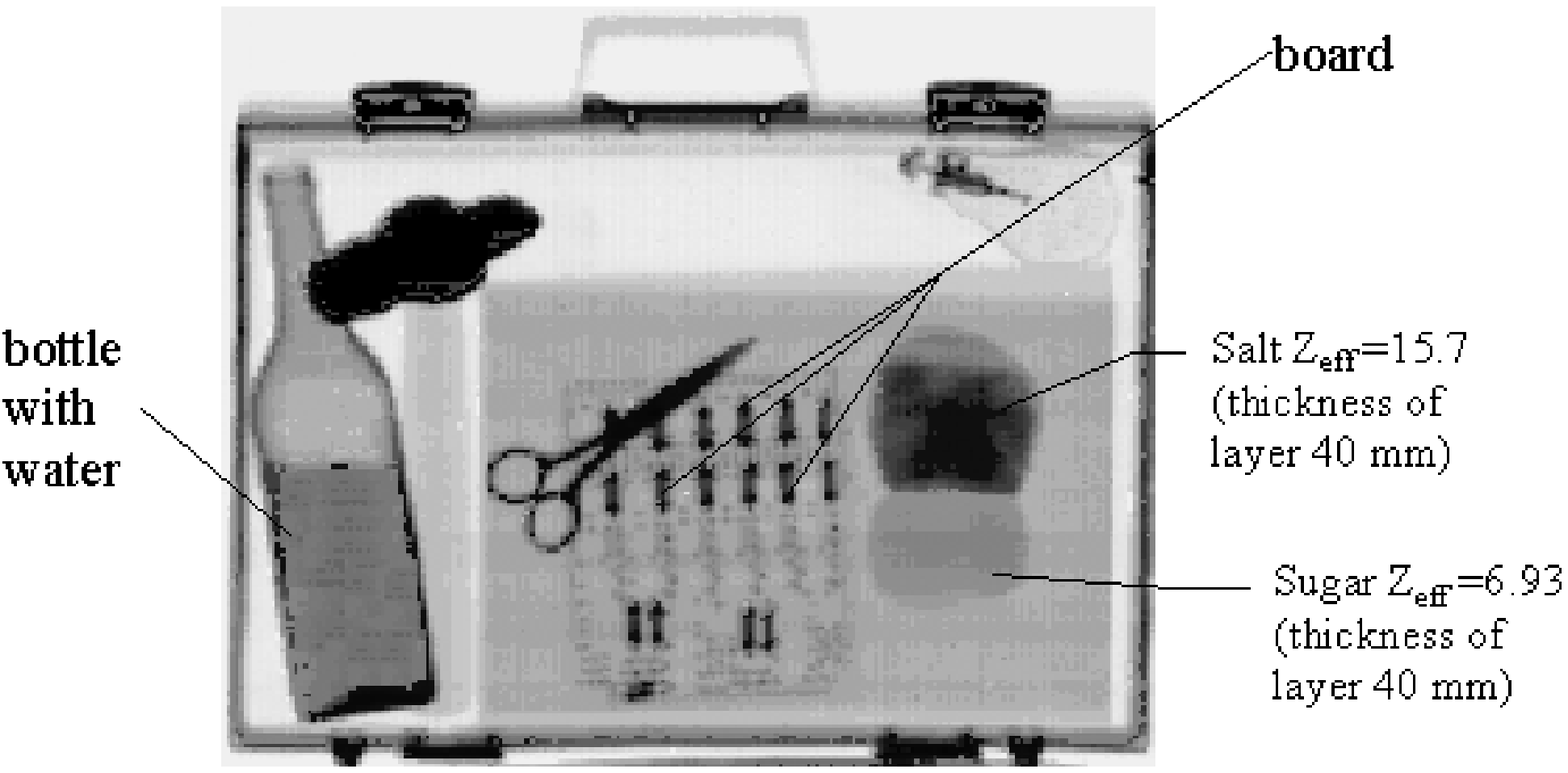}
\end{center}
\label{fig:Fig5a}
\end{figure}

\begin{figure}[h]
\begin{center}
\textbf{b)}
\includegraphics*[scale=1.2]{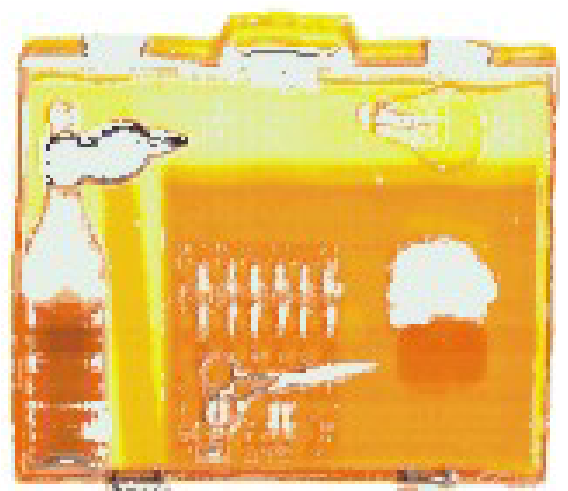}
\textbf{c)}
\includegraphics*[scale=1.2]{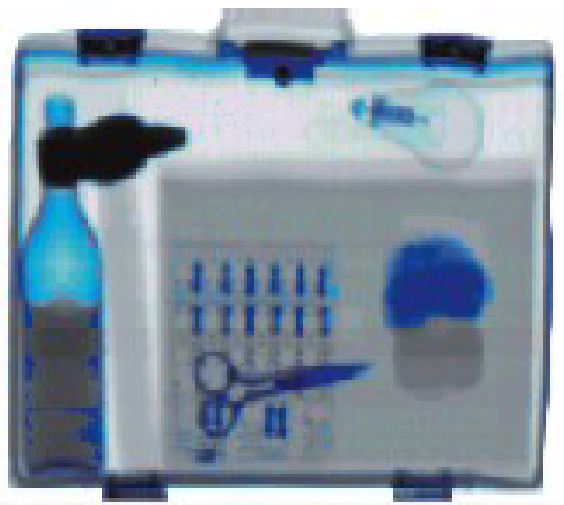}
\end{center}
\caption{Object images obtained using the two-energy introscope:
a) general shadow picture of the object; b) shadow picture with
inorganic material singled out; c) shadow image with organic
material singled out.} \label{fig:Fig5bc}
\end{figure}

\end{document}